\documentclass[review]{elsarticle}

\usepackage[margin=1in]{geometry}
\usepackage{float}
\usepackage{lineno,hyperref}
\usepackage{color}
\usepackage{amssymb}
\usepackage{graphics}
\usepackage{graphicx}
\usepackage{pbox}
\usepackage{mhchem}
\usepackage{xcolor} 
\usepackage[ampersand]{easylist}
\usepackage{lineno}
\usepackage{booktabs}
\usepackage{multirow}
\usepackage{subcaption}
\usepackage{amsmath}
\usepackage{paralist}
\usepackage{soul}
\usepackage{siunitx}
\usepackage{amsmath}
\usepackage{url}
\usepackage{makecell}
\usepackage{hyperref}
\usepackage{enumitem}

\newcommand{\cm}{\text{cm}} 
\newcommand{\g}{\text{g}}

\newcommand{\eff}{\text{eff}}

\newcommand{\PE}{\text{PE}}
\newcommand{\CS}{\text{CS}}
\newcommand{\PP}{\text{PP}}

\hypersetup{  
    colorlinks=true,
    linkcolor=blue,
    filecolor=magenta,      
    urlcolor=cyan,
    pdftitle={Sharelatex Example},
    bookmarks=true,
    pdfpagemode=FullScreen,
    citecolor=blue
    }

\modulolinenumbers[1]

\makeatletter
\def\ps@pprintTitle{%
 \let\@oddhead\@empty
 \let\@evenhead\@empty
 \def\@oddfoot{\centerline{\thepage}}%
 \let\@evenfoot\@oddfoot}
\makeatother

\begin{document}

\begin{frontmatter}

\title{Atomic number estimation of dual energy cargo radiographs: initial experimental results using a semiempirical transparency model}

\author[PNNLaddress,MITaddress]{Peter Lalor\corref{corauthor}}
\cortext[corauthor]{Corresponding author 
\\\hspace*{13pt} Email address: peter.lalor@pnnl.gov
\\\hspace*{13pt} Telephone: (925) 453-1876}

\author[MITaddress]{Areg Danagoulian}

\address[PNNLaddress]{Pacific Northwest National Laboratory, Richland, WA 99354, USA}
\address[MITaddress]{Department of Nuclear Science and Engineering, Massachusetts Institute of Technology, Cambridge, MA 02139, USA}

\begin{abstract}
To combat the risk of nuclear smuggling, radiography systems are deployed at ports to scan cargo containers for concealed illicit materials. Dual energy radiography systems enable a rough elemental analysis of cargo containers due to the Z-dependence of photon attenuation, allowing for improved material detection. This work presents our initial experimental findings using a novel approach to predict the atomic number of dual energy images of a loaded cargo container.  Our past work introduces a semiempirical transparency model, which is able to correct for bulk scattering effects, source energy uncertainty, and detector response uncertainty through a simple calibration procedure. The semiempirical model is more accurate than a fully analytic model, and shows improved extrapolation accuracy compared to existing empirical methods. This work considers measurements taken by a Rapiscan Sentry\textsuperscript{\textregistered} Portal scanner, which is a dual energy betatron-based system used to inspect cargo containers and large vehicles. We demonstrate the ability to accurately fit our model to a set of calibration measurements. We then use the calibrated model to reconstruct the atomic number of an unknown material by minimizing the chi-squared error between the measured pixel values and the model predictions. We apply this methodology to two experimental scans of a loaded cargo container. First, we incorporate an image segmentation routine to group clusters of pixels into larger, roughly homogeneous objects. By considering groups of pixels, the subsequent atomic number reconstruction step produces a lower noise result. We demonstrate the ability to accurately reconstruct the atomic number of blocks of steel and high density polyethylene. Furthermore, we are able to identify the presence of two high-Z lead test objects, even when embedded within lower-Z organic shielding. These results demonstrate the significant potential of this methodology to yield improved performance characteristics over existing methods when applied to commercial dual energy systems.
\end{abstract}
\begin{keyword}
Dual energy radiography \sep non-intrusive inspection  \sep atomic number discrimination \sep nuclear security
\end{keyword}
\end{frontmatter}
\begin{sloppypar}

\section{Introduction}
\label{Introduction}

Customs and Border Protection (CBP) processes more than 33.4 million imported cargo containers through U.S. ports of entry each year~\cite{CBP2022}. To combat the threat of nuclear terrorism, the U.S. Congress passed the Security and Accountability For Every (SAFE) Port Act of 2006~\cite{PLAW109-347}. The SAFE Port Act mandated 100 percent screening of U.S.-bound cargo and 100 percent scanning of high-risk containers as a means of improving port security.

Approximately 5 percent of seaborne containers are identified as high risk and scanned using non-intrusive inspection (NII) technology~\cite{CBO2016}. These radiography systems measure the attenuation of X-rays and/or gamma rays which are directed through the container to produce a density image of the scanned cargo. Some radiography systems deploy dual energy photon beams, enabling classification of objects according to their $Z$, since the attenuation of photons depends on the atomic number of the intervening material~\cite{Novikov1999, Ogorodnikov2002}. This technology improves the capabilities of these systems to identify nuclear threats or high-$Z$ shielding.

Our past work presented a novel method for predicting the area density and atomic number of dual energy radiographic images using a semiempirical transparency model~\cite{Lalor2023, lalor2024_estimation}. In this analysis, we apply these methods to a set of calibration scans taken by an AS\&E\textsuperscript{\textregistered} Rapiscan Sentry\textsuperscript{\textregistered} Portal scanner. The results of this study show the potential for these techniques to accurately distinguish between organic materials, inorganic materials, and heavy metals on experimental images taken by a commercial scanner.

\section{Background}
\label{Background}

When a radiography system scans a material of area density $\lambda$ and atomic number $Z$, it measures the transparency of the photon beam, defined as the detected charge in a scintillator-based sensor in the presence of the material normalized by the open beam measurement. Using the Beer-Lambert law, we define a free-streaming photon transparency model $T(\lambda, Z)$ as follows:

\begin{equation}
T(\lambda, Z) = \frac{\int_0^{\infty} D(E) \phi(E) e^{-\mu (E, Z) \lambda} dE}{\int_0^{\infty}D(E) \phi(E) dE}.
\label{transparency}
\end{equation}

In Eq.~\ref{transparency}, $\mu(E, Z)$ is the mass attenuation coefficient, $\phi(E)$ is the differential photon beam spectrum, and $D(E)$ is the detector response function. For this work, we calculate the photon beam spectrum and detector response function from the output of Geant4 simulations~\cite{Geant4, grasshopper}. The radiographic transparency is dependent on the atomic number of the imaged material through the mass attenuation coefficient $\mu(E, Z)$. Thus, by making multiple transparency measurements of the same object using different photon energy spectra, properties of the material $Z$ can be inferred. For the remainder of this analysis, we use the subscripts $\{H,~L\}$ to distinguish between the \{high, low\} energy beam measurements. Our past work found that the accuracy of Eq.~\ref{transparency} can be substantially improved by defining a semiempirical mass attenuation coefficient $\tilde \mu(E, Z)$~\cite{Lalor2023}:

\begin{equation}
\tilde \mu(E, Z) = a\mu_\PE(E, Z) + b\mu_\CS(E, Z) + c\mu_\PP(E, Z)
\label{semiempirical_mass_atten}
\end{equation}

where $a$, $b$, and $c$ are determined through a least-squares calibration step. In Eq. \ref{semiempirical_mass_atten}, $\mu_\PE(E, Z)$, $\mu_\CS(E, Z)$, and $\mu_\PP(E, Z)$ are the mass attenuation coefficients from the photoelectric effect (PE), Compton scattering (CS), and pair production (PP), calculated from NIST cross section tables~\cite{NIST}. We refer to this improved model (by substituting Eq. \ref{semiempirical_mass_atten} into Eq. \ref{transparency}) as the semiempirical transparency model.

\section{Methodology}
\label{Methodology}

We summarize the steps to apply these methods to a commercial system as follows:

\begin{enumerate}[label=(\arabic*)]
\item Calculate an approximate model of the system's beam energy spectra $\phi_H(E)$ and $\phi_L(E)$, and detector response function $D(E)$, using either Monte Carlo simulations or experimental measurements.
\item Perform at least three experimental calibration measurements, recording the true $\lambda$ and $Z$ of the calibration objects along with the measured transparencies $T_H$ and $T_L$. 
\item Calculate the calibration parameter values $a$, $b$, and $c$ which best reproduce the calibration measurements through a least-squares routine. $a$, $b$, and $c$ are functions of the detector channel index and should be calculated separately for the high- and low- energy beams.
\item To compute the pixel-by-pixel $Z_\eff$ estimate of a radiographic image, first perform an image segmentation step to group similar pixels. Then, for each pixel segment, perform a chi-squared minimization to find the optimal value of $Z$.
\end{enumerate}

Steps (1), (2), and (3) only need to be performed once. Steps (2) and (3) should be repeated if the system needs to be recalibrated, but do not need to be repeated between different container scans. A significant advantage of this approach is the simplicity of the calibration step, requiring only three calibration scans. Step (4) is then performed independently on each radiographic image. We emphasize that implementing this methodology does not require a detailed simulated model of the scanning system. For full algorithmic details and a validation using Geant4 Monte Carlo, see Ref.~\cite{lalor2024_estimation}.

\section{Analysis}
\label{Analysis}

\subsection{Calibration results}

\begin{figure}
\begin{centering}
\includegraphics[width=0.49\textwidth]{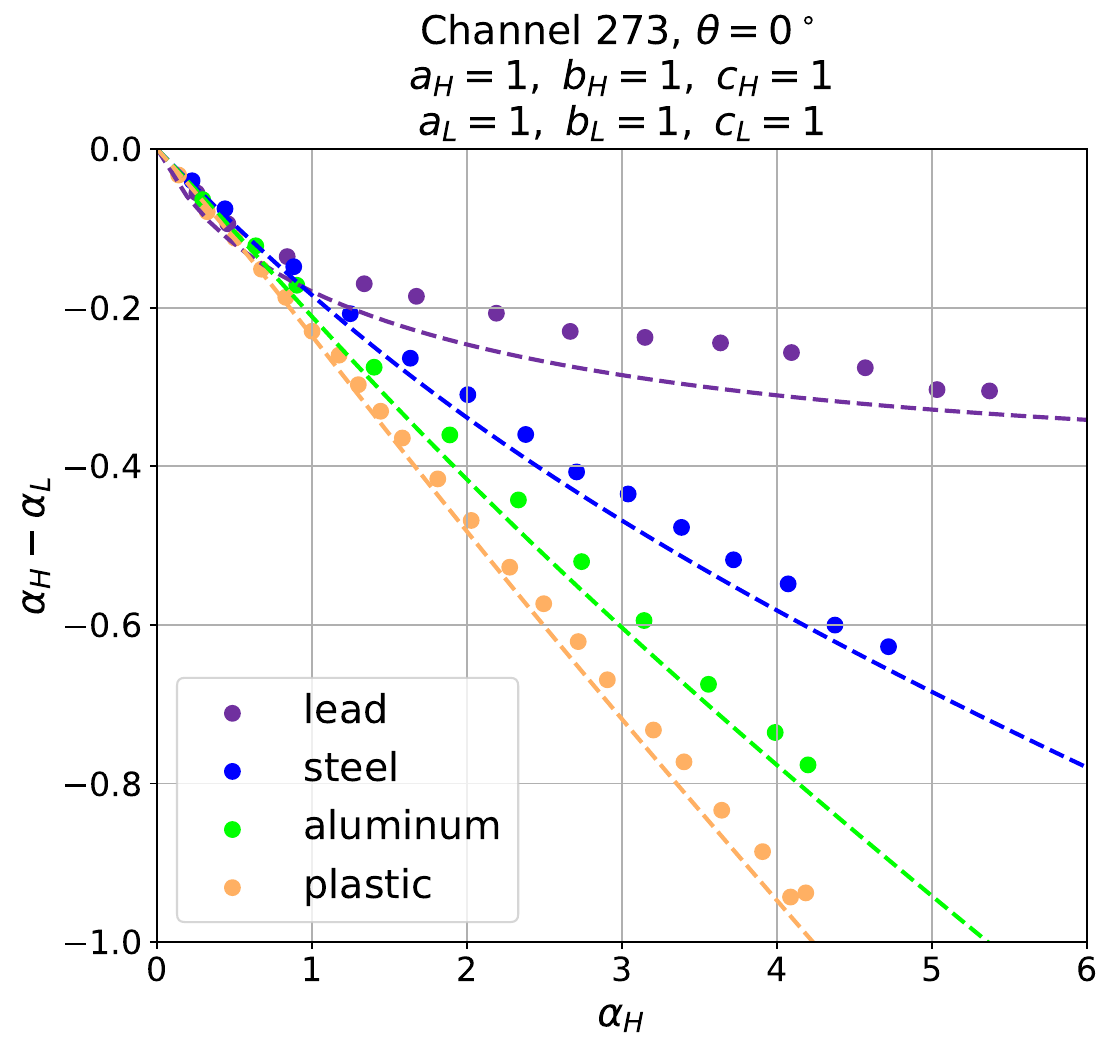}
\includegraphics[width=0.49\textwidth]{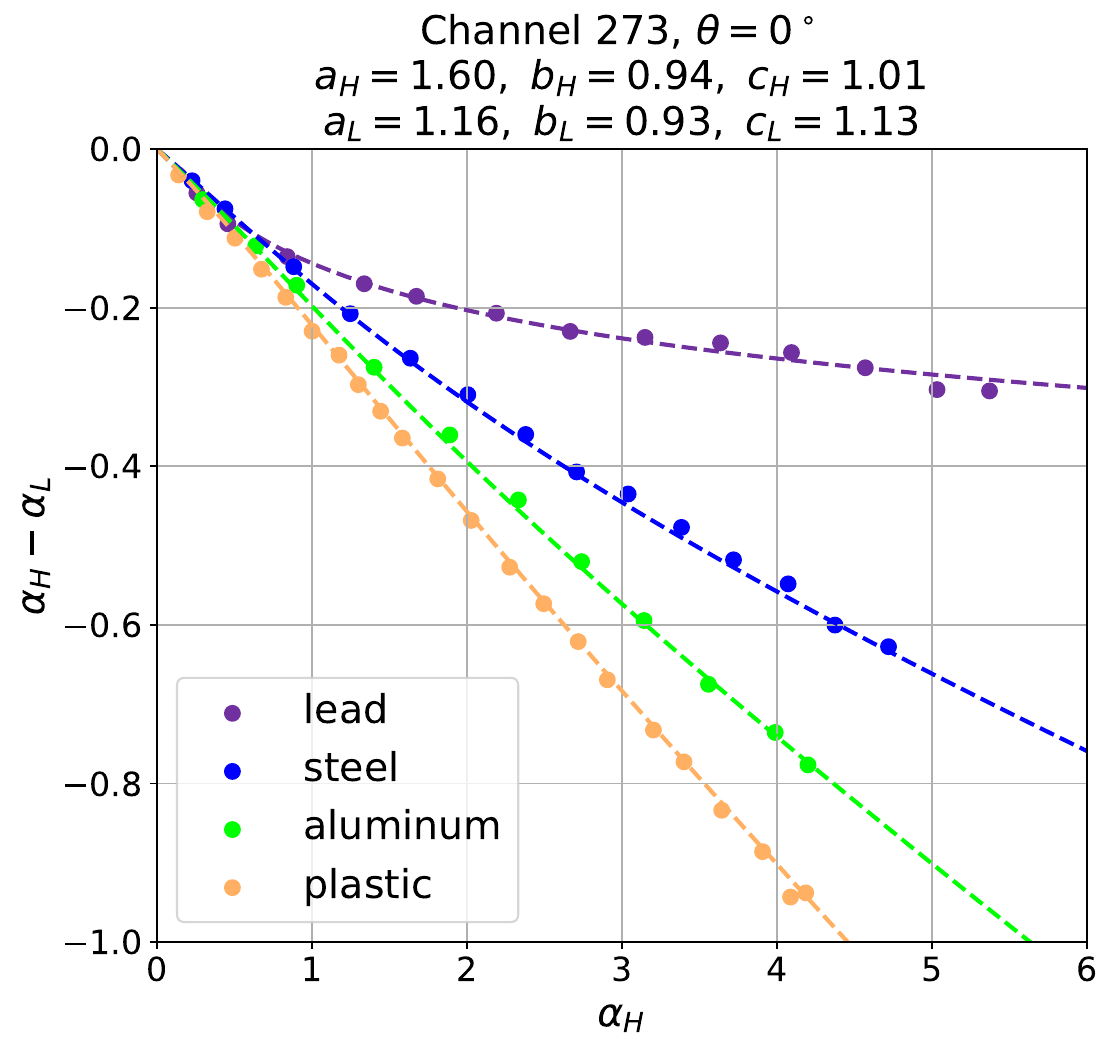}
\includegraphics[width=0.49\textwidth]{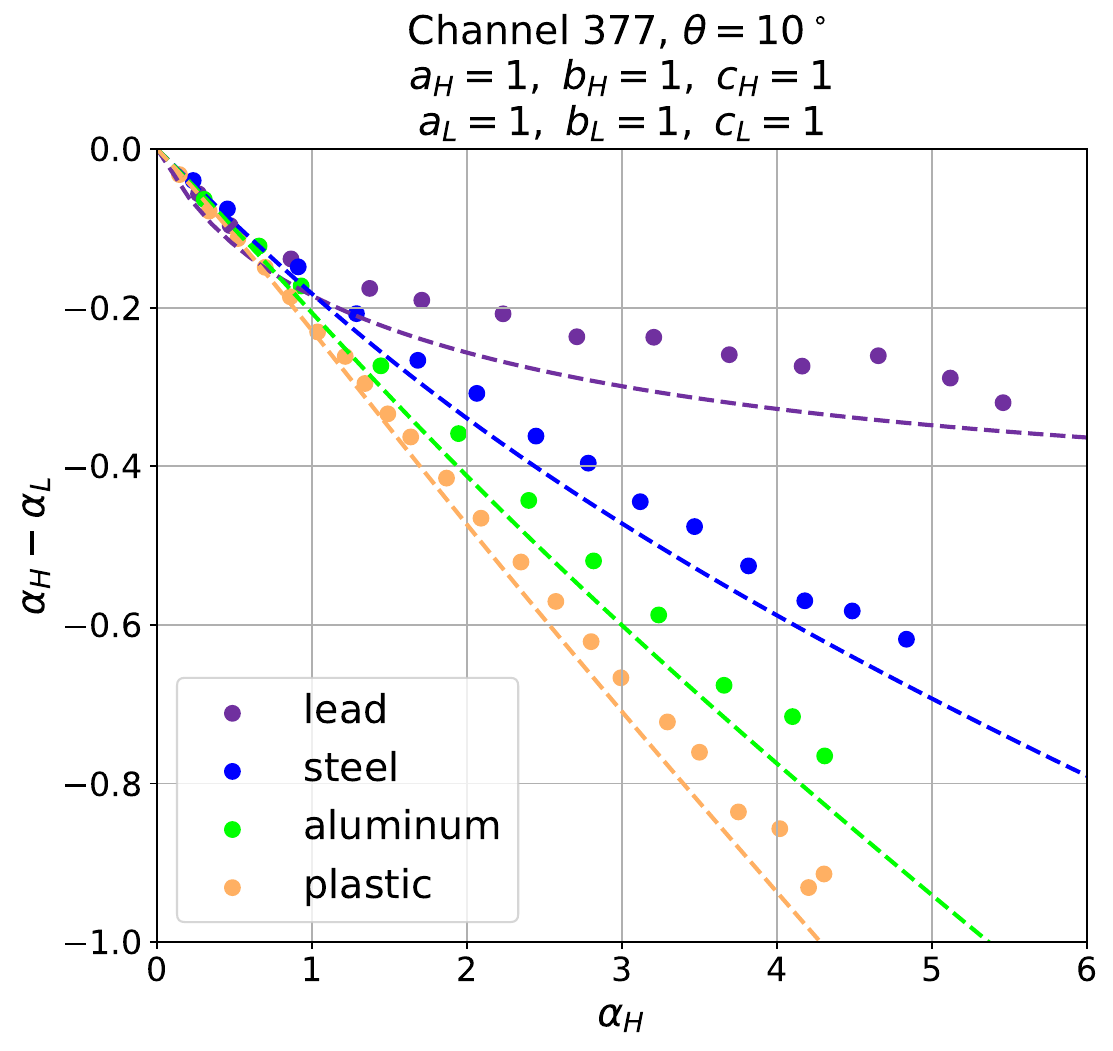}
\includegraphics[width=0.49\textwidth]{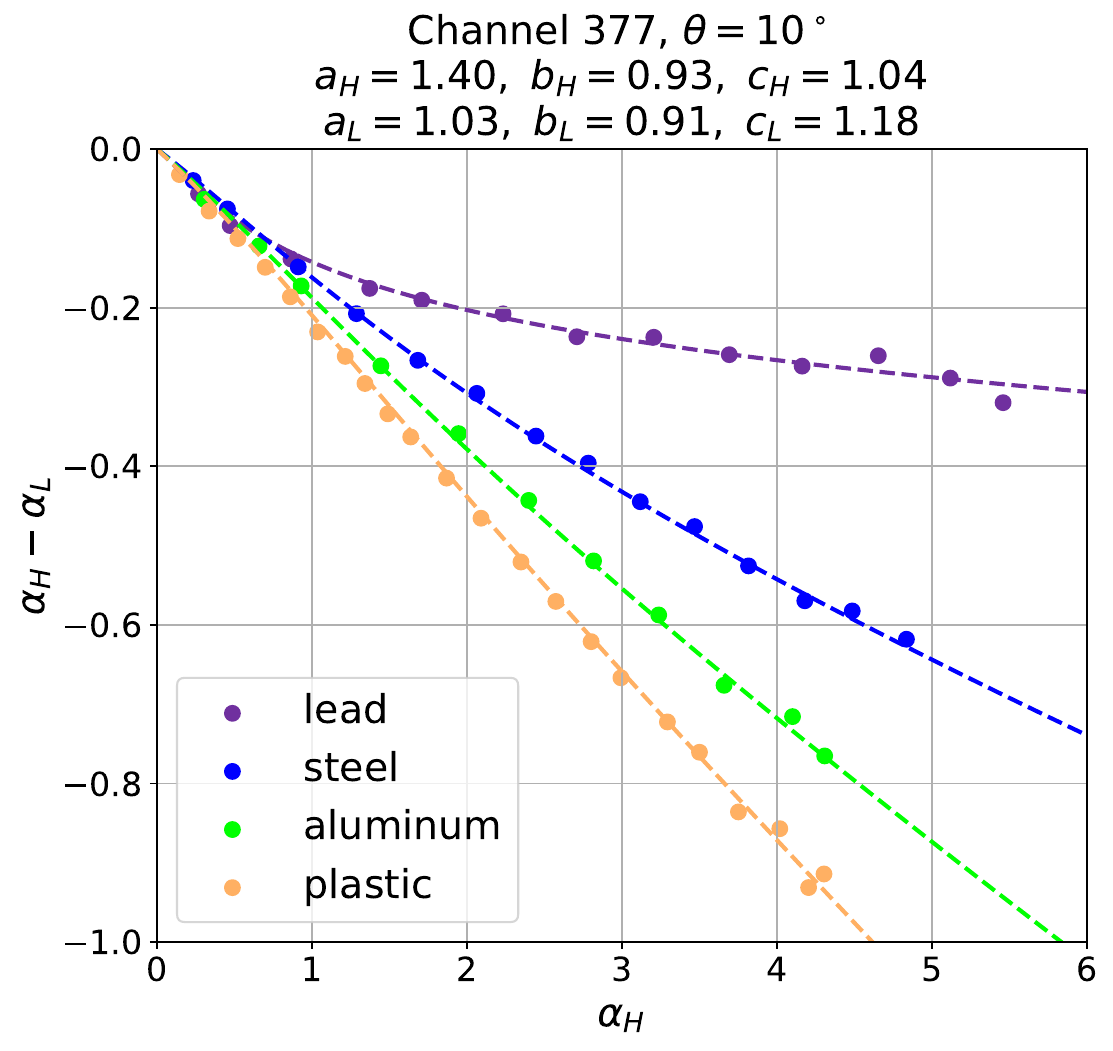}
\caption{Comparing experimental measurements (dots) to model predictions (dashed lines). The left plots show the uncalibrated free-streaming model, and the right plots show the calibrated semiempirical model. We include comparisons for two different detector channels at different beam angles.}
\label{fig:calibration_comparison}
\end{centering}
\end{figure}

In this analysis, we will consider experimental measurements taken by a Rapiscan Sentry\textsuperscript{\textregistered} Portal scanner. To assess the accuracy of the free-streaming and semiempirical models, we compare the model predictions to the calibration measurements on an $\alpha$-curve. An $\alpha$-curve is a plot of $\alpha_H - \alpha_L$ versus $\alpha_H$ for different materials and thicknesses, where we have performed a log transform $\alpha \rightarrow -\log T$. Every material on an $\alpha$-curve forms a characteristic $\alpha$-line, and separation between different $\alpha$-lines helps visualize material discrimination capabilities. In Fig. \ref{fig:calibration_comparison}, we see that the free-streaming model yields a consistently poor fit, while the semiempirical model shows excellent agreement with the calibration data. As the beam angle increases, the quality of the calibration data worsens, but the semiempirical model maintains a more accurate fit.

\subsection{Cargo Image Results}
\label{cargo image results}

\begin{figure}
\begin{centering}
\includegraphics[width=\textwidth]{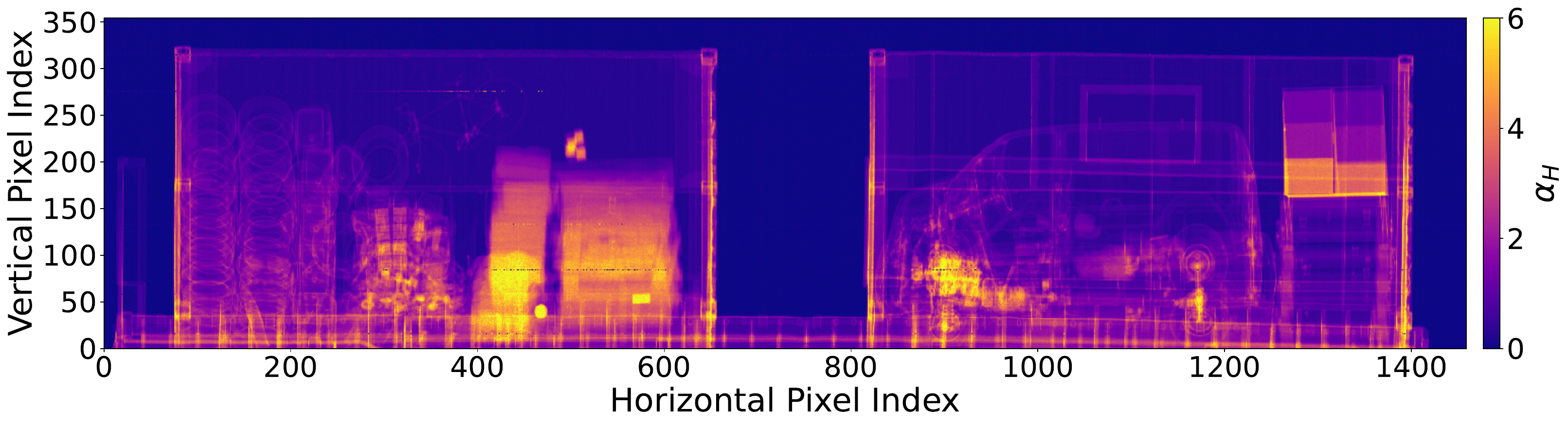}
\includegraphics[width=\textwidth]{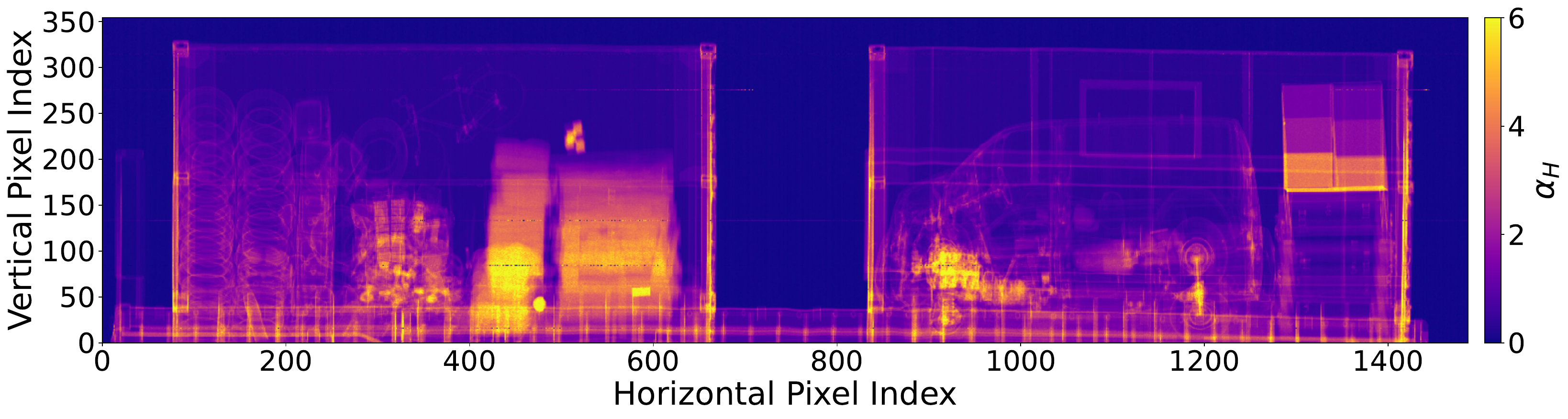}
\caption{Raw high energy scan 173 (top) and scan 174 (bottom) after preprocessing data.}
\label{fig:raw_scans}
\end{centering}
\end{figure}

\begin{figure}
\begin{centering}
\includegraphics[width=\textwidth]{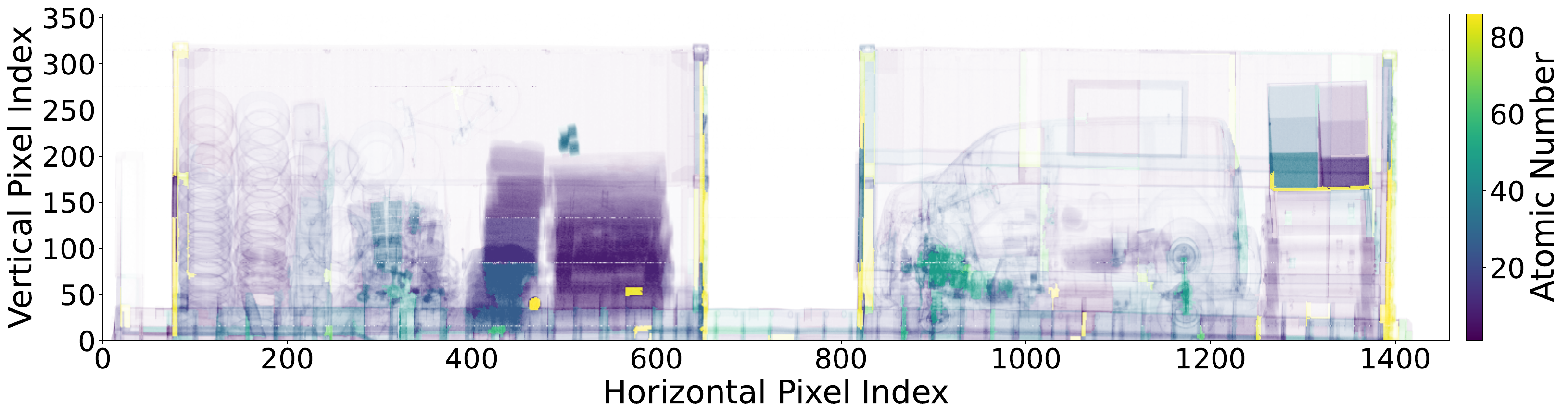}
\includegraphics[width=\textwidth]{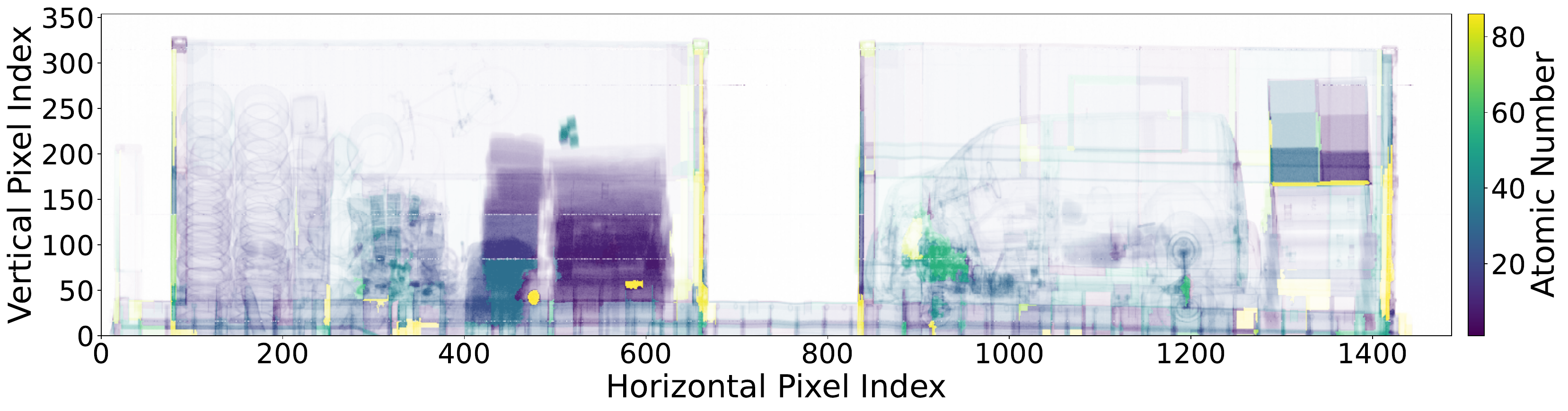}
\caption{Reconstructed atomic number map of scan 173 (top) and scan 174 (bottom). These results are quantified in Table~\ref{table:table_labels}. The algorithm is able to clearly distinguish between organic materials and inorganic materials.}
\label{fig:with_segmentation}
\end{centering}
\end{figure}

Using the methods of Section~\ref{Methodology}, we reconstruct the pixel-by-pixel atomic number and area density estimates of two separate dual energy images of the same container. We show the raw high energy scans in Fig.~\ref{fig:raw_scans}, and we show the resulting atomic number estimates in Fig.~\ref{fig:with_segmentation}. The results show strong overall agreement with the known material composition, although we do notice several high-$Z$ artifacts. These artifacts are likely a result of the fact that the container is moving, which means that the high- and low-energy beams do not pass through the exact same vertical strips of the container. This effect is especially important near vertical material boundaries, since the high-energy pulse might be attenuated while the subsequent low-energy pulse misses the object (or vice versa). Additionally, if the bremsstrahlung beam passes along the vertical boundary of an object, it is possible that only part of the beam is attenuated while the edge of the beam does not pass through the material. Both of these compounding effects result in unpredictable behavior near vertical material boundaries, so it is not surprising that the associated atomic number estimates deteriorate significantly. Ref.~\cite{Fu2010} gives a detailed discussion of the ``edge effect'' as a common source of false classification. For commercial applications, it will be necessary to correct for or otherwise filter out this signal corruption that occurs near vertical material boundaries.

\begin{figure}[t]
\begin{centering}
\includegraphics[width=\textwidth]{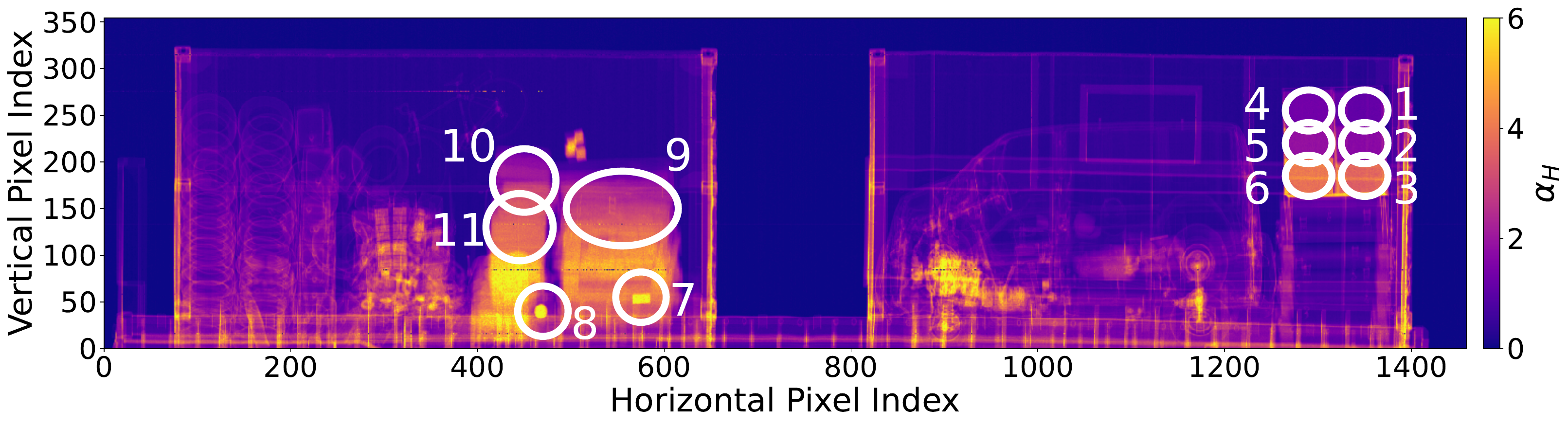}
\caption{Selected labels for inspection, along with their numbering. See table~\ref{table:table_labels} for the corresponding atomic number estimates of each region.}
\label{fig:selected_labels}
\end{centering}
\end{figure}

To quantify the accuracy of the atomic number reconstruction routine, we identify specific regions within the cargo container containing materials of known composition, as shown in Fig.~\ref{fig:selected_labels}. In Table~\ref{table:table_labels}, we show the atomic number estimate for each corresponding segment, along with the $25^\text{th}-75^\text{th}$ percentile area density estimate range. We see that we are able to easily identify high density polyethylene (HDPE) as organic, steel as inorganic, and lead as a heavy metal. Although there is a clear discrepancy between the true material $Z$ and the reconstructed $Z_\eff$, we observe overall strong agreement with the known material composition.

\begin{table}
\begin{centering}
\setlength{\tabcolsep}{1.5em}
\begin{tabular}{c c c c c c }
\toprule
  & & \multicolumn{2}{c}{Scan 173} & \multicolumn{2}{c}{Scan 174} \\
 Region & Material & $Z_\eff$ & $\lambda_\eff ~(\g/\cm^2)$ & $Z_\eff$ & $\lambda_\eff ~(\g/\cm^2)$\\
\midrule
 1 & HDPE & 5 & $27-29$ & 9 & $28-45$ \\
 2 & HDPE & 8 & $41-43$ & 9 & $28-45$ \\
 3 & HDPE & 11 & $104-115$ & 11 & $103-115$ \\
 4 & Steel & 28 & $28-30$ & 31 & $31-32$ \\
 5 & Steel & 28 & $42-44$ & 33 & $46-48$ \\
 6 & Steel & 35 & $102-110$ & 30 & $102-110$ \\
 7 & Lead brick & 84 & $118-126$ & 87 & $120-128$ \\
 8 & Lead pig & 86 & $118-128$ & 86 & $117-132$ \\
 9 & Coal & 9 & $46-94$ & 11 & $47-84$ \\
 10 & Rice & 9 & $46-94$ & 11 & $47-98$  \\
 11 & Rice & 13 & $108-147$ & 16 & $103-147$  \\
\bottomrule
\end{tabular}
\caption{Atomic number and area density estimates of the different regions in Fig.~\ref{fig:selected_labels}. We see strong agreement with the true material, with single digit atomic number errors and correct classification by material type (organic, inorganic, heavy metal).}
\label{table:table_labels}
\end{centering}
\end{table}

\section{Conclusion}
\label{Conclusion}

This analysis applies the semiempirical transparency model to experimental images taken by a commercial AS\&E\textsuperscript{\textregistered} Rapiscan Sentry\textsuperscript{\textregistered} Portal scanner. Using only a basic computational estimate of the scanner's beam spectra and detector response, we demonstrate that the semiempirical transparency model is able to obtain strong agreement with a set of calibration measurements. Furthermore, we are able to accurately distinguish between different types of materials (organics, inorganics, heavy metals) within an experimental test image. We incorporate a simple image segmentation step, which significantly reduces the variance of the calculated atomic number estimate. This result shows the potential for the methods described in this study to be applied to a range of different commercial scanners.

\section{Acknowledgements}

A portion of the research described in this paper was conducted under the Laboratory Directed Research and Development Program at Pacific Northwest National Laboratory, a multiprogram national laboratory operated by Battelle for the U.S. Department of Energy. Peter Lalor is grateful for the support of the Linus Pauling Distinguished Postdoctoral Fellowship program, and of the Department of Energy Computational Science Graduate Fellowship under grant DE-SC0020347. The authors would like to acknowledge Cristian Dinca at Rapiscan Systems for his useful suggestions and feedback.
\newpage

\bibliography{References.bib}

\end{sloppypar}
\end{document}